# Intergranular Hotspots: A Molecular Dynamics Study on the Influence of Compressive and Shear Work

Brenden W. Hamilton[1,2], Matthew P. Kroonblawd[3], Jalen Macatangay[1], H. Keo Springer[3], and Alejandro Strachan[1]*

Affiliations

[1]School of Materials Engineering and Birck Nanotechnology Center, Purdue University, West Lafayette, Indiana, 47907 USA
[2]Theoretical Division, Los Alamos National Laboratory, Los Alamos, New Mexico 87545, USA
[3]Physical and Life Sciences Directorate, Lawrence Livermore National Laboratory, Livermore, California 94550, USA

* strachan@purdue.edu

## Abstract

Numerous crystal- and microstructural-level mechanisms are at play in the formation of hotspots, which are known to govern high explosive initiation behavior. Most of these mechanisms, including pore collapse, interfacial friction, and shear banding, involve both compressive and shear work done within the material and have thus far remained difficult to separate. We assess hotspots formed at shocked crystal-crystal interfaces using quasi-1D molecular dynamics simulations that isolate effects due to compression and shear. Two high explosive materials are considered (TATB and PETN) that exhibit distinctly different levels of molecular conformational flexibility and crystal packing anisotropy. Temperature and intra-molecular strain energy localization in the hotspot is assessed through parametric variation of the crystal orientation and two velocity components that respectively modulate compression and shear work. The resulting hotspots are found to be highly localized to a region within 5-20 nm of the crystal-crystal interface. Compressive work plays a considerably larger role in localizing temperature and intra-molecular strain energy for both materials and all crystal orientations considered. Shear induces a moderate increase in energy localization relative to unsheared cases only for relatively weak compressive shock pressures of approximately 10 GPa. These results help isolate and rank the relative importance of hotspot generation mechanisms and are anticipated to guide the treatment of crystal-crystal interfaces in coarse-grained models of polycrystalline high explosive materials.

# 1. Introduction

Shockwaves in solids can induce a variety of complex responses such as plasticity[1–3], melting[4,5], fracture[6,7], and chemical reactions[8–10]. The shock initiation of chemistry, which can eventually lead to a run to detonation in high explosives (HEs), is governed by the formation of hotspots, which are local regions of excess temperature[11,12]. These pockets of high energy density can be formed through many mechanisms such as shear bands, cracking, friction, void collapse, and the interaction of shock waves[12]. Shock desensitization experiments have shown that the collapse of voids and porosity is the dominant mechanism behind the formation of critical hotspots that can result in the run to detonation[13]. Direct numerical simulations performed across a range of time and length scales have helped unravel the governing physics of hotspot formation and how that links to initial material microstructure[14,15,24,16–23], yet a complete mapping between initial microstructure and hotspot formation remains a grand challenge for the shock physics community.

Atomistic scale simulations have shown that, for strong shocks, the pressure-volume (PV) work done via recompression of ejected material during pore collapse is a key mechanism for reaching high temperatures that result in prompt chemistry[25]. As a shockwave reaches the upstream face of the pore, material accelerates and expands into the void. Once the expanded material propagates across the void, it recompresses as it collides with the downstream face of the pore, leading to excess PV work compared to the shock in bulk material, forming a hotspot. The more material can expand in the void, the more work will be done during recompression. Shock focusing and compression along the major axis of high-aspect-ratio defects can eject molecules into the void to a gas-like density, which maximizes the PV work done during recompression[26].

Pore geometry can have substantial effects on the collapse process. With 2D and 3D void geometries (i.e., surfaces/faces with curvature), shock focusing can create a laterally inwards flow of material to form a jet[27,28]. For simple 1D cases or a planar surface, material accelerates to twice the shock's particle velocity when the wave reaches a free surface. Such 1D "voids" serve as a reductionist model for the expansion and compression along the minor axis of high-aspect-ratio pores and for intergranular geometries encountered in real HE samples, which are almost always polycrystalline in nature. Holian et. al. derived an expression for the maximum increase in temperature, $\Delta T_{\max}$, for hotspots formed in the 1D case, where PV work is the main mechanism due to a lack of jetting and limited effects from friction and shear[25]. The scaling law for this maximum is: $k_B \Delta T_{\max} = m U_s U_p / d$ where $m$ is atomic mass, $d$ is dimensionality, $U_s$ and $U_p$ are the shock and particle velocities, and $k_B$ is Boltzmann's constant.

Recent molecular dynamics (MD) studies have shown that while high temperatures trigger prompt reactions within a hotspot, an additional mechanism can further accelerate and alter these reactions: mechanochemistry resulting from latent intra-molecular potential energy. Mechanochemistry, or chemistry that results from the straining and deformation of covalent bonds, has been well documented for a variety of systems[29–34]. Reactive MD simulations have shown that hotspots formed during the shock compression of voids are significantly more reactive than those formed in the perfect crystal under equivalent temperature and pressure conditions.[35] Adding a shear component during planar crystal-crystal impact (or '1D pore collapse') can also increase reactivity within the hotspot[36].

Nonreactive MD modeling has shown that plastic flow during pore collapse in the molecular crystalline HE TATB (1,3,5-trimamino-2,4,6-trinitrobenzene) can lead to large intra-molecular strains in which the strain energy is stored directly in the modes relevant to prompt chemistry[37,38]. Reactive calculations of the same intra-molecular strain phenomena showed that these strains both accelerate initial reactions and alter the first step reaction pathways[39,40]. Nanoscale shear banding in bulk TATB was shown to also induce large intra-molecular strains which leads to a significant acceleration of kinetics[41] that occurs essentially homogeneously throughout the material for shocks above approximately 20 GPa[42]. A steered MD approach designed to capture the complex many-body observed in these strained molecular states was able to systematically extract reaction kinetics and paths, showing up to 25% decrease in activation energy for deformations found in hotspots and the possibility of alternate reaction pathways that have a higher energy barrier under unstrained conditions[43]. Previous works have shown these molecular strains and mechanochemical response to be driven by plastic flow during the formation of the hotspot[38], and that by varying shock strength, pore size, and crystal orientation in TATB, different levels of molecular strain were induced.

The above prior work indicates the importance of expansion and recompression to generate high temperatures in hotspots, as well as localized high-rate plastic flow to deform molecules leading to their "mechanochemical activation". Shock-induced pore collapse exhibits both processes in a highly coupled manner and typically involve complex geometries, which makes it challenging to separate their contributions to energy localization and the initiation of chemistry.

To address this challenge, we use MD simulations specially designed to independently control expansion/recompression and shear deformation. We adopt a quasi-1D simulation geometry in which two spatially separated crystals are subjected to a compressive shock with prescribed lateral load that induces shearing at the crystal-crystal interface formed upon impact. By varying the longitudinal and transverse loads as well as the crystallographic orientation of the HE crystals, we map the processes that localize energy and find conditions that enhance intra-molecular strain and temperature of the hotspots that form at these interfaces. The generality of the trends is assessed by comparison of two HE materials that differ in their molecular shape and conformational flexibility.

## 2. Methods

All simulations in this work were performed with all-atom non-reactive MD using the LAMMPS software[44,45]. Two representative HE materials were considered, TATB and pentaerythritol tetranitrate (PETN). Both materials were modeled using non-reactive force fields with similar class-I functional forms.

The force field used for TATB is based on that of Bedrov et al.[46], and includes tailored harmonic bond stretch and angle bend terms for flexible molecules[47], and an intra-molecular O-H repulsion term that was implemented as a bonded interaction[48]. The covalent bond vibrations, angle bends, and improper dihedrals are modeled using harmonic functions. Proper dihedrals are modeled using a cosine series. Van der Waals interactions are modeled using the Buckingham potential (exponential repulsion and a $r^{-6}$ attractive term) combined with short-ranged $r^{-12}$ potentials that compensate for the divergence in the Buckingham potential at small separation. The non-bonded terms were evaluated in real space within an 11 Å cutoff. Electrostatic interactions were calculated between constant partial charges located on the nuclei and were evaluated using the short-ranged Wolf potential with a damping parameter of 0.2 Å$^{-1}$ and an 11 Å cutoff [49]. The TATB force field excludes all intra-molecular nonbonded interactions by design.

The force field used for PETN was developed by Borodin et. al.[50]. We used the same implementation of the PETN force field as described in Ref. [51]. There are three key differences between the TATB and PETN force field forms and implementation in LAMMPS. First, the PETN force field does not include short-ranged r$^{-12}$ potentials that compensate for the divergence in the Buckingham potential. Second, it employs standard intra-molecular nonbonded exclusions in which only the 1-2 and 1-3 nonbonded interactions are set to zero. Third, electrostatics in the PETN force field were evaluated with the PPPM method[51,52] with the relative accuracy set to $1 \times 10^{-6}$ rather than with the Wolf potential. We note that while our prior testing has shown negligible differences in atomic forces obtained with the Wolf potential and PPPM for TATB,[41] the present simulations were not sufficiently large to motivate testing the Wolf potential as applied to the PETN force field. Implementing the PETN force field using the Wolf potential would provide significant computational speedup for larger systems.

Our simulation approach was designed to assess the localization of energy at crystal-crystal interfaces subject to axial compression and transverse shearing. Simulation cells were modeled after the work in Ref. [36] in which a 1D void (or gap) was placed at the center of the cell between two slabs of crystal with periodic boundary conditions in the lateral directions, as shown in Figure 1. The sample (both crystals) is launched into a fixed wall by assigning a particle velocity to all atoms, shear loading is controlled by a lateral velocity applied to one of the crystals as described in detail below. Gaps with a length of ~40nm, were created using multiples of whole unit cells, with initial cell sizes ~250 nm in length along the compression direction and ranging from 4-6nm along the transverse directions. Cells were thermalized at 300 K using a Nose-Hoover thermostat[53] for 100 ps. Atomic velocities were re-initialized every 10 ps to attenuate breathing modes that form upon the creation of a free surface.

Shock simulations were performed with the reverse ballistic approach[54] by holding the leftmost 5nm of the sample rigid, which forms a piston that drives a shock wave in the remaining fully flexible portion of the sample. Two translational velocity components were added to the thermal velocities of the flexible molecules in each crystal to impose an initial condition leading to compression and shear. Compressive work was controlled by adding the shock particle velocity (i.e., compressive velocity) $U_p = V_z$ along the z direction to both crystals. Shearing work was controlled by adding a second lateral velocity $U_\tau = V_y$ to the second crystal on the right-hand side along the y direction. This leads to a situation where the second crystal moves laterally at a uniform velocity until complete closure of the void space when the two crystals impact each other leading to shear friction at the interface.

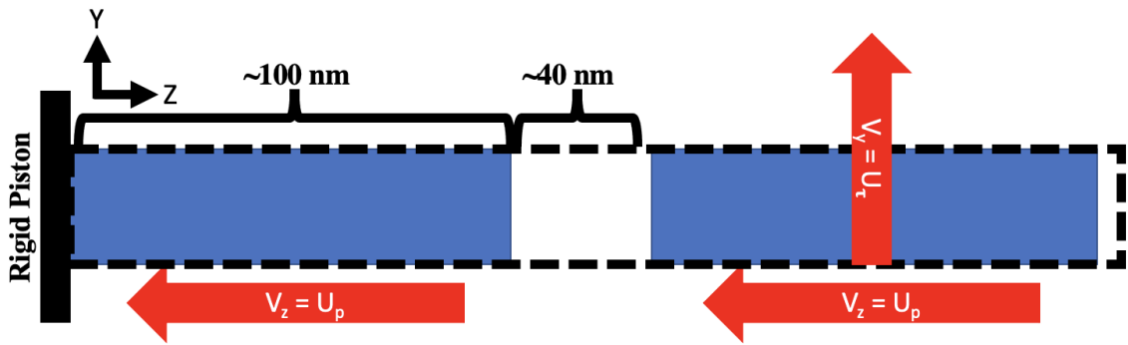

*Figure 1: Simulations setups for crystal – crystal shock impacts with varying compressive and lateral velocities. Dashed line represents periodic boundary conditions.*

Three crystallographic orientations were considered for TATB, and two for PETN, as shown in Figure 2. Constructions for both HEs used equilibrium lattice parameters determined with their respective force fields at 1 atm and 300 K, with the TATB parameters coming from Ref. [48] and the PETN parameters from Ref. [51]. The TATB oriented cells were produced with the generalized crystal cutting method (GCCM)[55]. TATB orientations were defined by the inclination angle $\theta$ of the shock direction vector **S**, which varied between the [100] direction ($\theta = 0°$) and the basal plane normal vector given by $\mathbf{a} \times \mathbf{b}$ ($\theta = 90°$). The GCCM solutions used to obtain oriented TATB supercells from the triclinic $P\bar{1}$ structure were as follows:

TATB 0°:
$$\mathbf{A} = -1\mathbf{a} - 2\mathbf{b} - 1\mathbf{c}$$
$$\mathbf{B} = 0\mathbf{a} + 0\mathbf{b} - 1\mathbf{c}$$
$$\mathbf{C} = \mathbf{a} + 0\mathbf{b} + 0\mathbf{c}$$

TATB 45°:
$$\mathbf{A} = -1\mathbf{a} + 1\mathbf{b} - 2\mathbf{c}$$
$$\mathbf{B} = 0\mathbf{a} + 3\mathbf{b} + 4\mathbf{c}$$
$$\mathbf{C} = 15\mathbf{a} + 11\mathbf{b} + 0\mathbf{c}$$

TATB 90°:
$$\mathbf{A} = -5\mathbf{a} - 3\mathbf{b} + 0\mathbf{c}$$
$$\mathbf{B} = \mathbf{a} - 7\mathbf{b} + 0\mathbf{c}$$
$$\mathbf{C} = \mathbf{a} + 2\mathbf{b} + 6\mathbf{c}$$

The supercells were oriented such that **A** was along x, **B** was in the x × y plane with a positive y component, and **C** was in the positive z half-space. The shock direction z, was therefore nominally along **C**, and the lateral velocity direction, y, was nominally along **B**. PETN orientations were chosen to span its observed[56,57] directional sensitivity to shock initiation, with [100] being more insensitive than [001]. The shear direction for both is the [010] direction. Noted orientations in Figure 2 are aligned with the vertical axis of the cell, which is the shock direction and is parallel to the z-axis arrow.

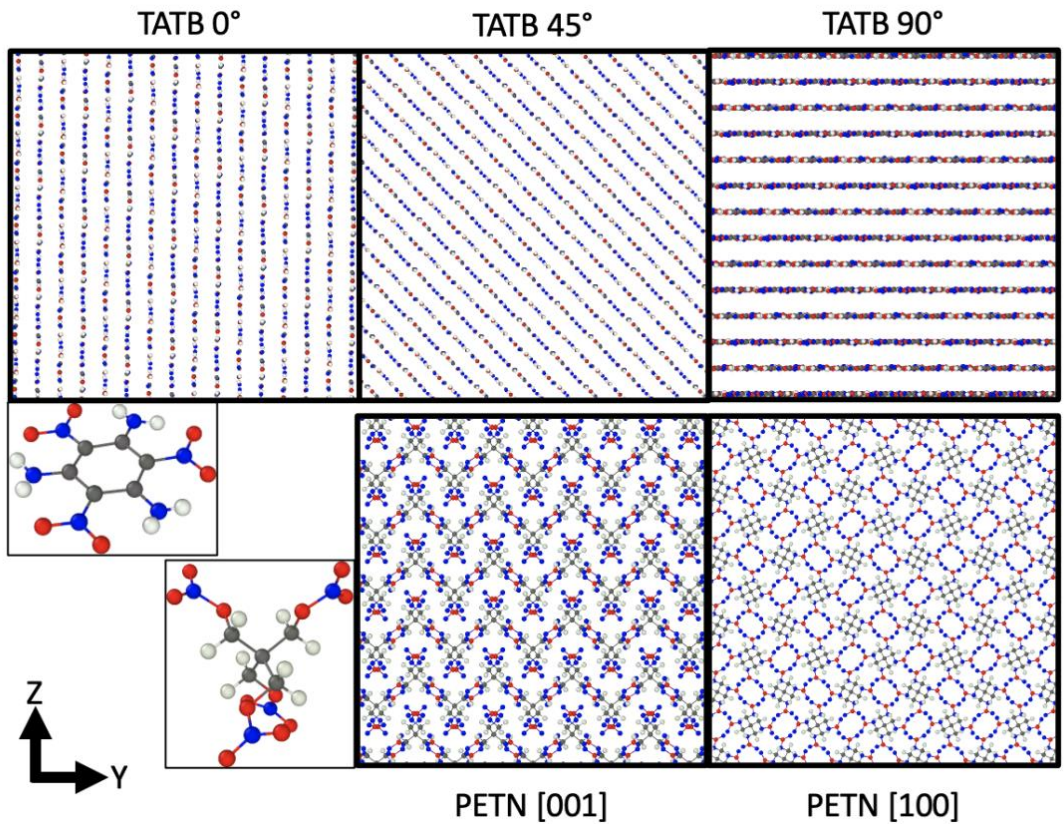

*Figure 2: Crystallographic structure of each of the oriented cells in relation to the compression (z) and shear (y) directions. Listed angles or crystallographic directions are aligned with the vertical (z) axis of the cell. Atoms are colored grey, blue, red, and white for C, N, O, and H.*

Two main properties were assessed on a per-molecule basis, the roto-vibrational kinetic energy ($K_{ro-vib}$) measured from the C.O.M. of each molecule, which we interpret as intra-molecular temperature ($T$) and use units of Kelvin, and the intra-molecular potential energy as calculated by the force field ($U_{Intra}$).

$$K_{ro-vib} = \frac{3N-3}{2} k_B T$$

$$U_{Intra} = \sum PE_{Bond} + \sum PE_{Ang} + \sum PE_{Dih} + \sum PE_{Imp} + \sum PE_{NB}$$

For TATB, the force field excludes all non-bonded intra-molecular interactions by design, so the intra-molecular potential energy can be directly computed from the sum of all bonded terms. For PETN, the force field includes intra-molecular non-bonded interactions, so the total per-molecule non-bonded energy contains both intra- and inter-molecular contributions in condensed-phase systems. We determined $U_{Intra}$ for PETN by calculating the total potential energy of each molecule through single-point calculations in which the molecule was placed in isolation in a large cubic cell, which effectively eliminates inter-molecular non-bonded contributions. The difference between $U_{Intra}$ and $K_{ro-vib}$ was also assessed, where each was referenced by the equilibrium value at 300 K and 1 atm. This difference gives a measure of the intra-molecular strain energy associated with deforming molecule conformations from the equilibrium structure, which we denote as $U_{Latent}$:

$$U_{Latent} = [U_{Intra} - U_o] - \left[\frac{3N-3}{2}k_B(T - 300 \text{ K})\right]$$

For an undeformed system in equilibrium, this value is zero on average due to the equipartition of energy. Supplemental Materials section SM-1 shows the $U_{Intra}$ increase in the bulk material for each compressive velocity applied.

## 3. General Shock Response

Figure 3 shows position-time diagrams (colloquially referred to as x-t diagrams) for a representative simulation with a 2.0 km/s compressive velocity and 3.0 km/s lateral velocity for the TATB 0° orientation case. These are colored by compressive direction (particle) velocity, the lateral (shear) velocity, and the local density, all of which are calculated using an Eulerian binning along the compression direction z with bins of 2nm. Figure 4 shows individual snapshots at key times in the same simulation as Figure 3. The view in Figure 4 captures material motions, and in particular highlights how the respective wave fronts and free surfaces coincide with phases of material expansion and recompression. We note that both the Figure 3 and 4 plots do not change qualitatively for different TATB or PETN orientations. Quantitative changes observed for these materials are discussed in Sections 4 and 5, respectively.

Focusing first on the compressive velocity in the left-most panel of Figure 3 shows the evolution of multiple wave fronts. The initial shockwave (a) transits the first crystal before reaching the first free surface. Once the wave reaches the surface (I), part is reflected back into the sample (b) while the material expands freely into the void (c). The expanding material impacts the second crystal at (II), which leads to a second reflection (d) and transmission of another supported shockwave (e). Comparison against the lateral velocity plot shows that even at high compression, the lateral velocity does not attenuate outside of the immediate crystal-crystal interface at II on picosecond timescales. Transverse material flow continues to occur well after crystal-crystal impact, a trend that significantly differs from previous 2D void collapse work[37,39]. This indicates relatively small friction in the shear band created around the impact plane. The density plot shows similar wave profiles as seen in the compressive velocity, but adds two additional pieces of information: (1) the material from the first crystal that expands into the void space (c) reaches a density state similar to the uncompressed crystal before it impacts the downstream surface a (II); and (2) the density is similar on either side of the downstream interface after impact at (II).

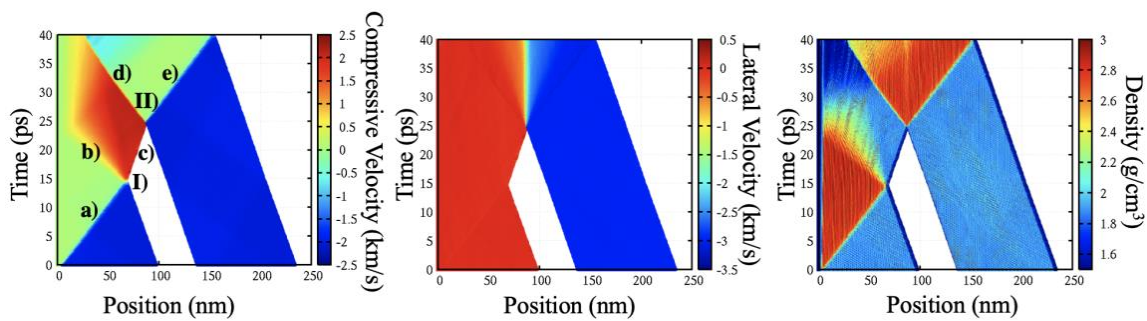

*Figure 3: x-t diagrams for the 2 km/s compressive velocity and 3.0 km/s lateral velocity, colored by compressive velocity, lateral velocity, and density.*

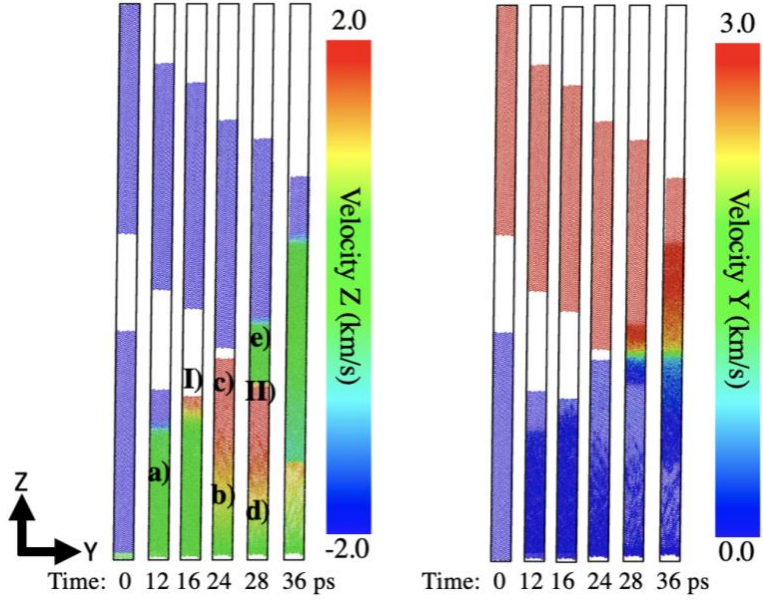

*Figure 4: Molecular renderings of the 0° case for Vz of 2.0 km/s and Vy of 3.0 km/s. Coloring is based on molecule center of mass velocities.*

## 4. TATB

We first focus on hotspots formed at crystal-crystal interfaces in TATB. A total of 105 distinct cases were considered in which the shock direction was set to values of $\theta = 0°, 45°$, and $90°$, the compressive velocity ranged from $V_z = 1$ km/s to 2 km/s in steps of 0.25 km/s, and the transverse velocity ranged from $V_y = 0$ km/s to 3 km/s by steps of 0.5 km/s. Figure 5 shows configuration snapshots that qualitatively highlight the orientation-dependent crystal deformations induced for a subset of these compressive and lateral velocities, denoted $V_z$ and $V_y$ respectively. Compressive velocities shown are 1.25 km/s and 2 km/s, lateral velocities shown are 0 and 3 km/s. These two compressive velocities lead to shocks with approximate pressures of 10 and 25 GPa, respectively, with the precise value being somewhat dependent on shock direction.

The 0° case results in significant disorder that varies with both $V_y$ and $V_z$. The crystal layers buckle, molecules rotate, and increasing shear velocity generally increases the degree of disorder at the interface for both weak and strong compressive shocks. Distinctly spaced planes where the layers buckle is more evident for the weaker compressive shock velocity. This is consistent with Ref. [58] which found that the spacing between buckling planes generally decreases with increasing strain rate.

In the 45° case, the crystal layers are oriented such that they align with the anticipated plane of maximum resolved shear at ±45° relative to the compression direction. This activates basal slip in which the crystal layers slide past each other without inducing more significant deformations that destroy local lattice packing. Similar basal glide was observed under shock and non-shock axial loads in Refs. [42,58,59]. For this orientation, the two cases with nonzero lateral velocity leads to modest disordering that is localized to the crystal-crystal interface.

The 90° case, which compresses normal to the crystal layers along the most compliant direction in the crystal[42], does not induce much disorder until both the compressive and lateral velocities reach high levels. Although difficult to discern with the image resolution, the crystal layers remain largely intact even in the case with $V_y$ = 3.0 km/s and $V_z$ = 2.0 km/s. Regions of disorder are observed behind the shock front and typically consume the entire width of the sample through the periodic boundary. This contrasts with the extensive nanoscale shear band network that has been shown to form for compression along this direction with much larger MD simulations[41,42]. Based on an earlier finite size effect study[41], it should be noted that the system cross-section sizes used here are expected to somewhat suppress the formation of shear bands in the bulk. This removes an important source for additional intra-molecular strain in the bulk far away from the interface. For non-overdriven shocks, the secondary plastic front where shear bands form will not reach the free surface before the primary elastic front causes the crystal to expand into the void space. Thus, we would not necessarily expect shear bands to form in the first crystal near the free surface with the present simulation geometry, even for cells with larger cross sections. Thus, we anticipate that the suppression of shear bands will not lead to qualitative differences in the structure of the hotspot formed at the crystal-crystal interface for this orientation case.

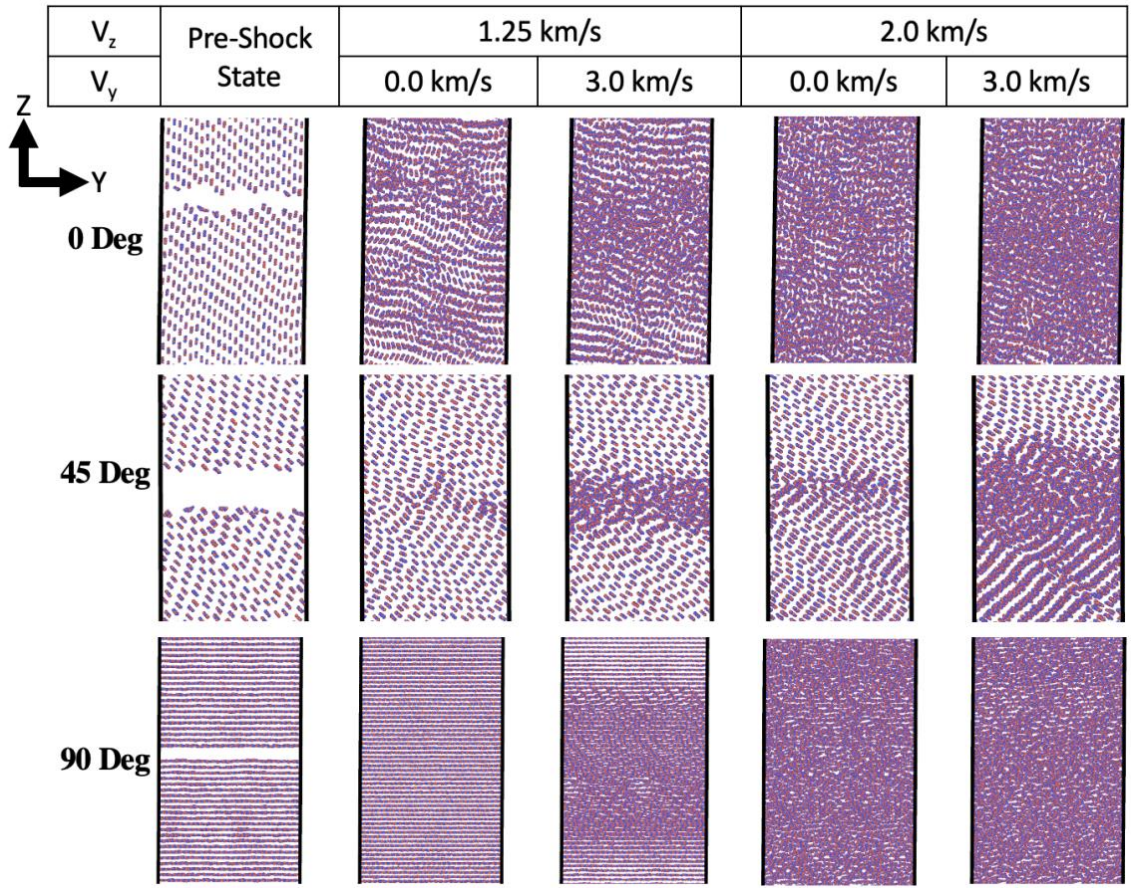

*Figure 5: Snapshots showing TATB packing structure within and around the hotspots formed at crystal-crystal interfaces for selected orientations, compressive velocities, and lateral velocities. Only C atoms in the TATB ring are visualized.*
9

Figure 6 contains x-t diagrams for all three orientations, showing the temperature and $U_{Latent}$ for the 2.0 km/s compressive velocity and 3.0 km/s lateral velocity case. Focusing first on the temperature, it can be seen in all cases that heating is largely uniform except on the upstream half of the crystal interface after impact of the two crystals ($x \approx 100$nm, $t \geq 25$ps). This is expected due to significantly more PV work done, as the material on the upstream surface expanded to near the initial density in the void space before it was recompressed on the downstream face of the second crystal. It is perhaps interesting that free expansion of the first crystal into the void leads to very little temperature dissipation. This contrasts with the response of $U_{Latent}$. While shock compression leads to large and positive $U_{Latent}$ in the first crystal, it dissipates to approximately zero upon free expansion into the void. Recompression leads to similar magnitude increase in $U_{Latent}$ that is largely symmetric across the crystal-crystal interface.

The most salient orientation-independent features of hotspots formed at crystal-crystal interfaces are the apparent asymmetry of the temperature field and symmetry of the intra-molecular strain energy field. Additional differences in the general responses of these two fields are also apparent. In these high shear velocity cases, a thin band of extreme temperature (>2500K) exists directly at the interface where the material has experienced large levels of friction. In terms of absolute magnitude, the average $U_{Intra}$ does not reach significantly higher values in the hotspot at the interface compared to its value in the compressed bulk.

Comparison of the three orientations shows that the hotspot temperature and bulk shock temperature are largely independent of orientation. Temperature in the bulk is governed by the input energy (compressive velocity) and plastic work, and the hotspot temperature is a function of the PV work done during expansion and recompression. However, for $U_{Latent}$, there is considerably more excess energy in the 0° case, with 90° being the lowest. This tracks well to the level of disorder shown in Figure 4, as the cases with the most apparent disorder result in the highest intra-molecular strain energy. Previous work on compression-induced energy localization in TATB[42] showed that the 0° case exhibits a moderately homogenous localization of intra-molecular strain energy driven by buckling/twinning defects in the crystal, which is also noted here. For cases in which $\theta \geq 45°$, shear banding begins to be the dominant deformation mechanism, which nucleates significantly more intra-molecular strain energy than the buckling/twinning mechanism. However, the small cell sizes in the lateral directions here suppress shear band formation, leading to significantly lower bulk energy localization than expected. While this does imply that orientation-dependent trends in $U_{Latent}$ should be interpreted carefully, we expect these finite size effects to have limited influence over the general features of the interfacial hotspot identified across the different orientations.

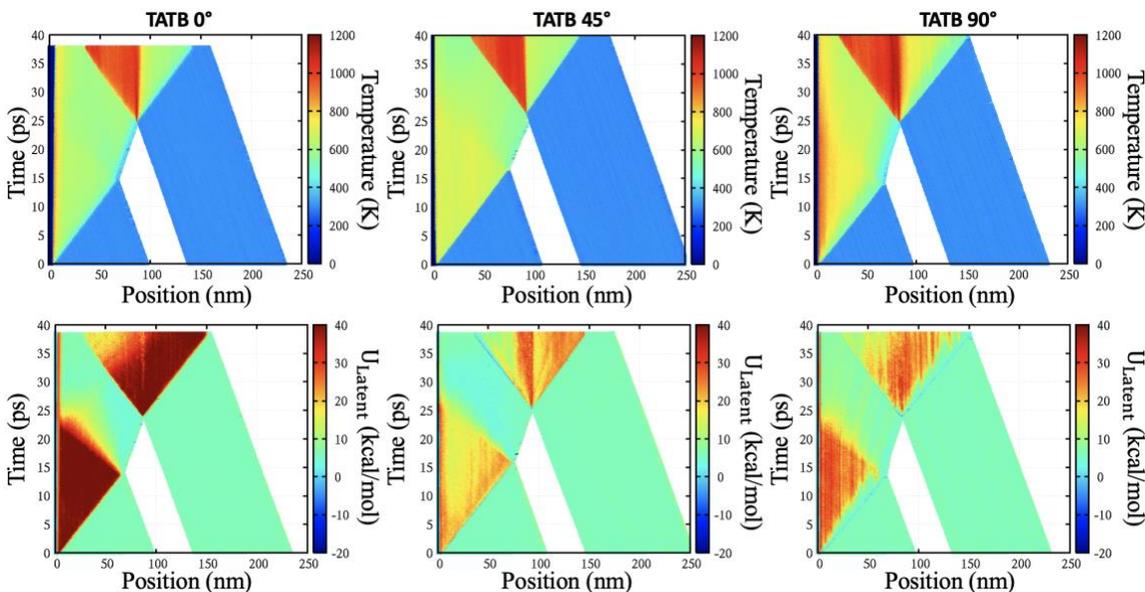

*Figure 6: x-t diagrams for the 2 km/s compressive and 3 km/s lateral velocity case for each orientation, colored by vibrational temperature and intra-molecular strain energy, $U_{Latent}$.*

Figure 7 shows distributions of $T$ and $U_{Intra}$ (note the distinction with $U_{Latent}$), where all molecules within the system are plotted 5ps after crystal-crystal impact. The cyan line represents the equipartition of energy, the expected value of $U_{Intra}$ if all potential energy is thermal (where $U_{Latent}$ is zero). Deviations from this line represent the $U_{Latent}$. Panel a) shows the 0° case with no lateral velocity, with increasing compressive velocity. With increasing compressive work, the temperature and $U_{Intra}$ increase, as expected, and an increase in the deviation from the equipartition line is also seen. Thus, it is clear that $U_{Latent}$ also increases with increasing shock strength. Panel b) shows the 2.0 km/s compressive velocity for various lateral velocities. This shows that increasing the lateral velocity increases the peak $T$ and $U_{Intra}$ reached, due to imparting more total energy into the system. However, by adding interfacial shear during recompression, there is not a noticeable difference in the deviation from equipartition. Root mean-square (RMS) deviations from the equipartition line for increasing lateral velocity at a constant 2.0 km/s compressive velocity are 35.32, 35.66, and 36.03 kcal/mol for 0.0, 1.5, and 3.0 km/s lateral velocity, respectively. Conversely, increasing the compressive velocity at a constant lateral velocity of 0.0 km/s gives RMS deviations of 27.96, 30.49, and 35.32 kcal/mol, for the 1.0, 1.5, and 2.0 km/s cases, respectively.

In previous work on TATB using a reactive force field,[39] we identified clear links between the reaction kinetics and both the molecular temperature and intra-molecular strain energy. Clustering analysis based on $T$ and $U_{Latent}$ identified distinct regimes where chemistry was accelerated due to mechanical strain of the molecules, which provides a measure of "mechanochemical activity". Thresholds for different levels of mechanochemical activity based on this clustering analysis are denoted by the orange and pink lines in Figure 7b, which respectively correspond to $U_{Latent}$ = 50 kcal/mol and 100 kcal/mol. Molecules above the pink line are in highly mechanochemically active states, and molecules below pink but above orange are moderately mechanochemically active.

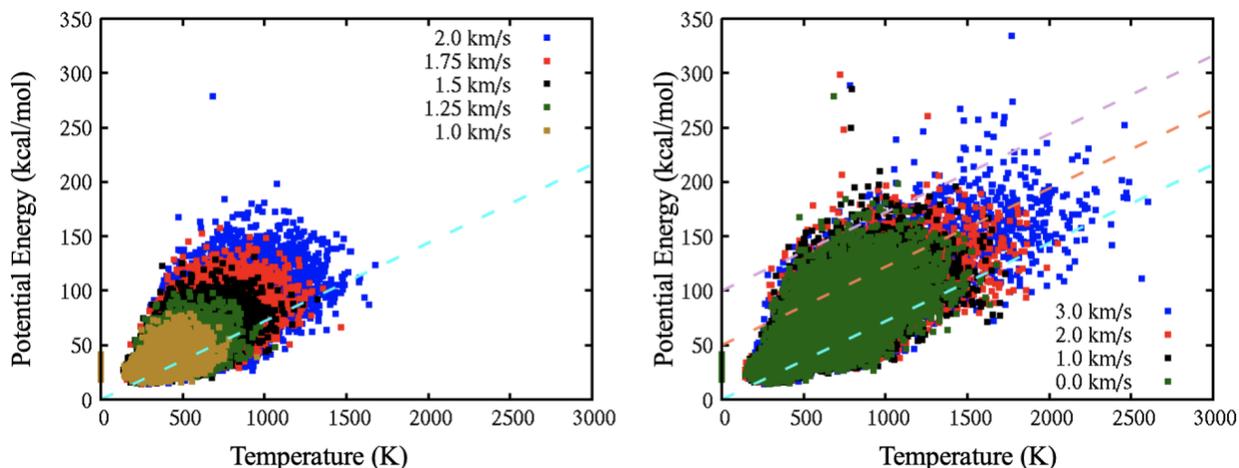

*Figure 7: PE-T ($U_{intra}$ and roto-vibrational T, respectively) distributions for the 0° TATB case. Panel a) shows increasing compressive velocity with no lateral velocity, panel b) shows increasing lateral velocity at 2.0 km/s compressive velocity. Cyan dashed lines represent perfect equipartition of energy. Purple and orange lines in panel b) represent regions of mechanochemical activity as defined in Ref. [39] where above the grey line corresponds to mechanochemically active (Reference Clusters 3+4 and 5+6).*

Figure 8 shows a 1D binning, with 2nm bins, of the population of the mechanochemical states along the compression direction for the 0° case and four sets of velocities that bookend our sampled ranges, i.e. compressive velocity of 1 and 2 km/s and lateral velocity of 0 and 3 km/s. For the 1 km/s compressive velocity (P ~10GPa) along the top row of Figure 8, the inclusion of a lateral velocity causes a thin spike of highly mechanochemical molecular states directly at the interface. These mechanochemical states are localized to molecules within a few unit cells of the surface as the z-dependent profiles show this region to be only 4-6nm thick. For 2 km/s compressive velocity (P ~25 GPa), there is very little effect from the lateral velocity, but the population of highly mechanochemical molecular states is significantly higher than for the weaker compressive velocity. It is perhaps surprising that lateral shearing has an almost negligible effect for this stronger compressive velocity. This implies that the compressive work is what creates this ~15nm thick region of almost all highly mechanochemical states. While the number of high $U_{Intra}$ outliers and the RMS deviation from equipartition in Figure 7b do not greatly increase with lateral velocity, it follows that lateral velocity does not impart a significant difference in the intra-molecular strain energy of the molecules in this mechanochemically activated interfacial region. We also note that there is an (artificially) enhanced mechanochemical activity near the piston at z = 0 nm compared to the bulk, especially for the cases with stronger compressive velocity.

Qualitatively similar trends to those just identified for the 0º case also hold for the 45º and 90º cases. We find that shear velocity enhances the population of mechanochemically activated molecules at the interface for weak compressive velocities, but not for strong compressive velocities. Figure 9 shows profiles of mechanochemical molecule populations for the weak and strong compressive velocities (top and bottom rows) at a lateral velocity of 3 km/s for the 45º and 90º cases in the left and right columns, respectively. These directly track the right-hand column in Figure 8 for the 0º case. A complete set of figures for the 45º and 90º cases analogous to Figure 8 can be found in the Supplemental Materials section SM-2.

For both orientations and compressive velocities, the left crystal, which expanded into the void space and was then recompressed, is made of predominately mechanochemical molecules, albeit at mostly "moderate" levels. In contrast, the downstream (right) crystal has mainly non-mechanochemical molecules with some moderately active mechanochemical states. All three orientations show a similar magnitude population of highly mechanochemical molecules in a 10-15nm interfacial region with a strong 2.0 km/s compressive velocity. While cross-inspection with the bottom row plots of $U_{Latent}$ in Figure 6 shows that there is symmetry of the average intra-molecular strain energy across the interface, the population analysis in Figures 8 and 9 indicates that the two crystals exhibit distinctly different distributions of molecular $U_{Latent}$ states on either side of the interface for all three orientations. A similar piston effect as was identified for the 0º case is also seen for the stronger compressive velocity with both the 45º and 90º orientations.

Orientation effects are most pronounced at the slower compressive velocity. The 0° and 45° cases exhibit a thin 4-6nm wide peak with ~80% of the molecules in this region being in highly mechanochemically activated states. The 90° case exhibits a similar peak at the interface, but with only ~25% of the molecules reaching high levels of mechanochemical activation. This is potentially due to the resilience of the hydrogen-bonded TATB crystal layers. The 90° case compresses directly perpendicular to the layers, with the lateral velocity running across them. These layers remain largely intact even during expansion and recompression of the first crystal, given the suppression of shear banding, which helps hold the molecules in a planar geometry and hence a low-$U_{Latent}$ state. Significantly more compressive velocity may be needed to break crystal layers and deform molecules at spatial scales below typical shear band dimensions, which are roughly 10nm wide and are spaced 10s of nm apart[41,42]. Comparison against Figure 5 shows that there is not much deformation of the layers at a compressive velocity of 1.25 km/s and a lateral velocity of 3.0 km/s. The degree of deformation increases substantially as the compressive velocity is increased to 2.0 km/s, which directly tracks with the enhanced population of mechanochemically activated molecules under those conditions seen in Figure 9.

Using the same cluster-based thresholds for counting mechanochemically activated molecules discussed above, we performed a comprehensive population analysis to identify quantitative trends with compressive and lateral velocity. Our population analysis considered a 50nm wide region centered on the crystal-crystal interface, which is shown in Figure 10 for the 0º case with each line corresponding to a different compressive velocity. It is immediately apparent that the total population of mechanochemically activated molecules is more dependent on compressive velocity than shear velocity, with almost no influence from shear velocity above a compressive velocity of 1.5 km/s. Similar qualitative trends follow for the 45° and 90° cases, which are shown in Supplemental Materials section SM-3. Pearson correlation coefficients for spatially resolved properties of the hotspot are also provided in Supplemental Materials section SM-4, which consider the cross-correlation of initial velocities, temperature, $U_{Latent}$, and the compressive and shear work. These corroborate the conclusions from Figures 7-10, showing that $U_{Latent}$ is highly correlated with the compressive velocity and work, but much less so with the lateral velocity and shear work.

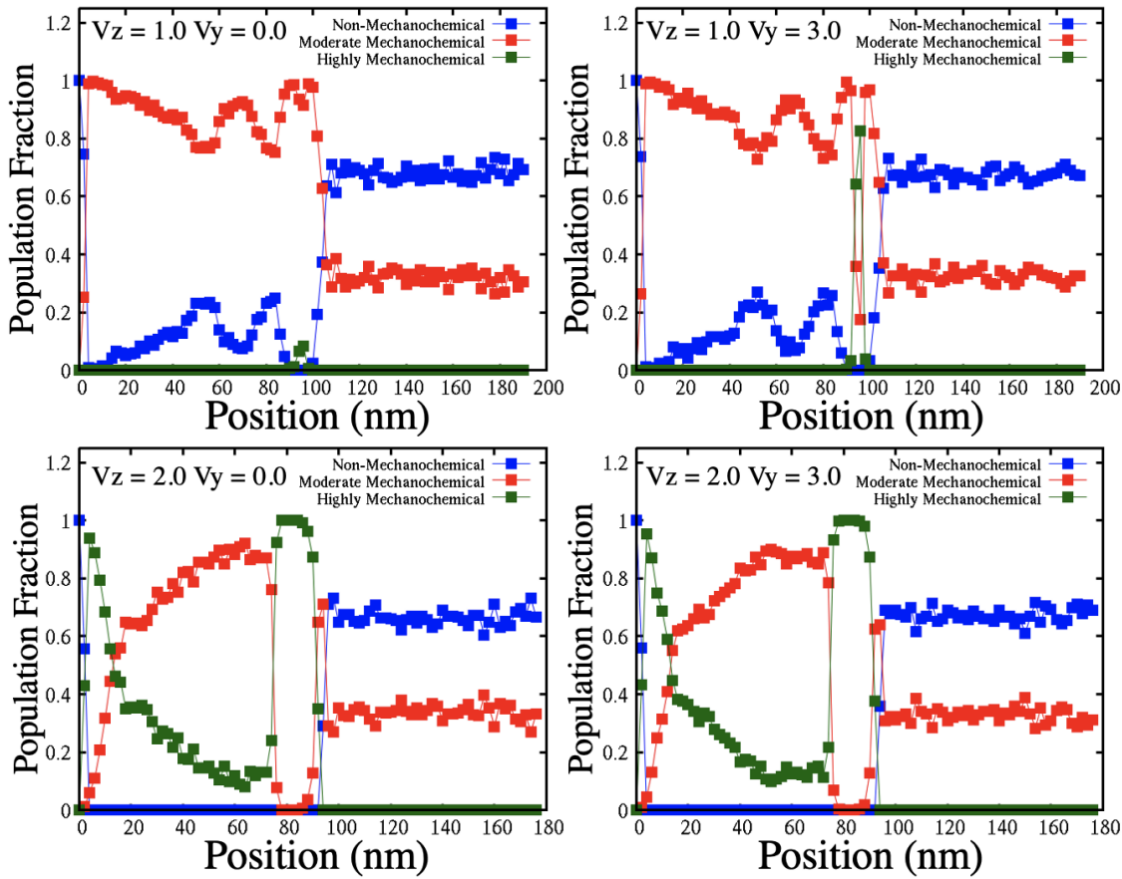

*Figure 8: Spatially resolved populations of mechanochemically activated molecules for the 0° TATB case at time $t_o + 5$ ps. Populations are based on the regions plotted in Figure 7b, which correspond to the mechanochemical model from Ref. [39]. Population fractions are computed with respect to the total number of molecules in each Eulerian bin. Panels correspond to 1.0 and 2.0 km/s compressive velocity, top and bottom row respectively, and 0.0 and 3.0 km/s lateral velocity, left and right columns, respectively.*

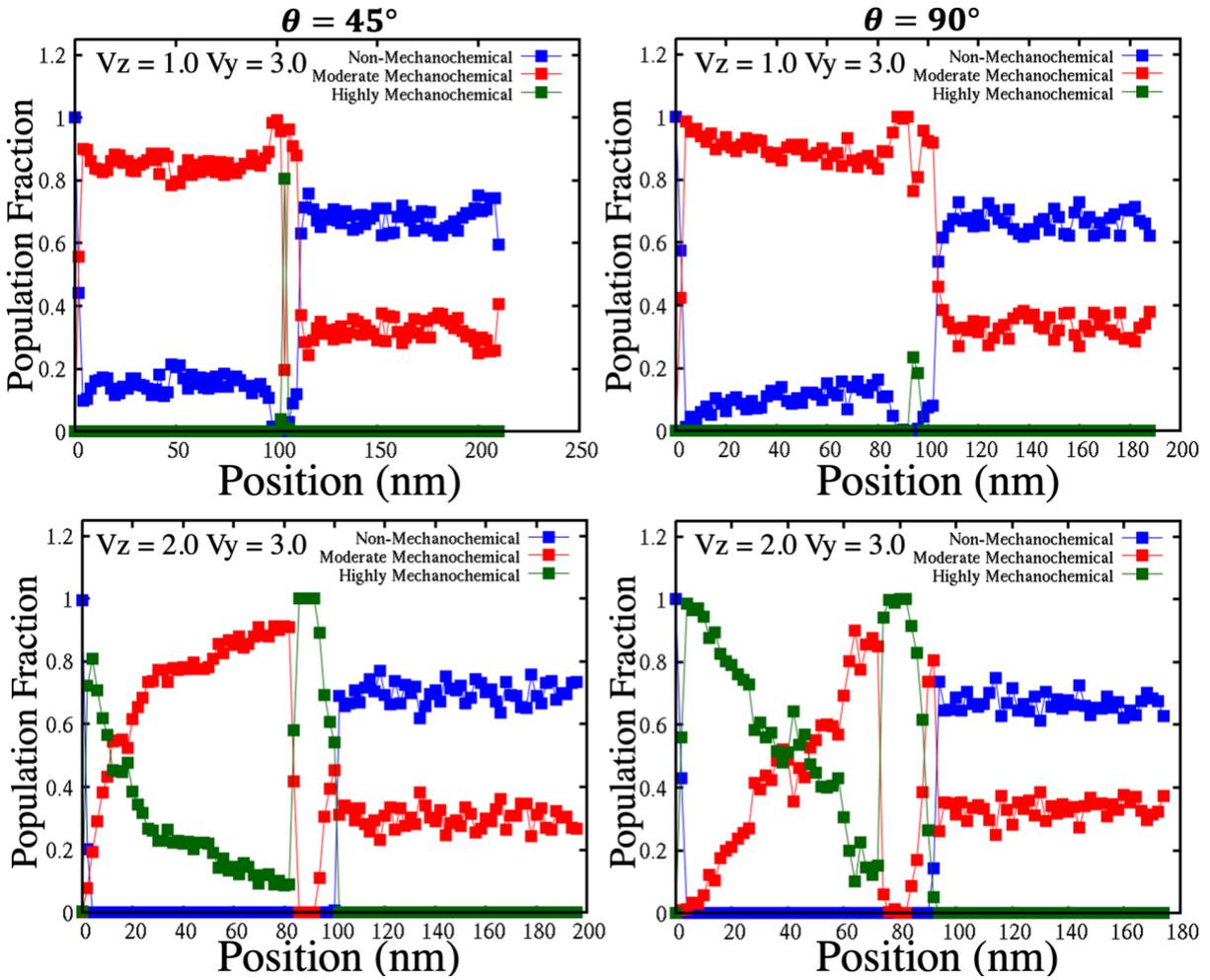

*Figure 9: Spatially resolved populations of mechanochemically activated molecules for the 45º and 90º TATB cases at time $t_o + 5$ ps. Populations are based on the regions plotted in Figure 7b, which correspond to the mechanochemical model from Ref. [39]. Population fractions are computed with respect to the total number of molecules in each Eulerian bin. Panels correspond to 1.0 and 2.0 km/s compressive velocity, top and bottom row respectively, at 3.0 km/s lateral velocity, with the left and right columns showing the 45º and 90º cases, respectively.*

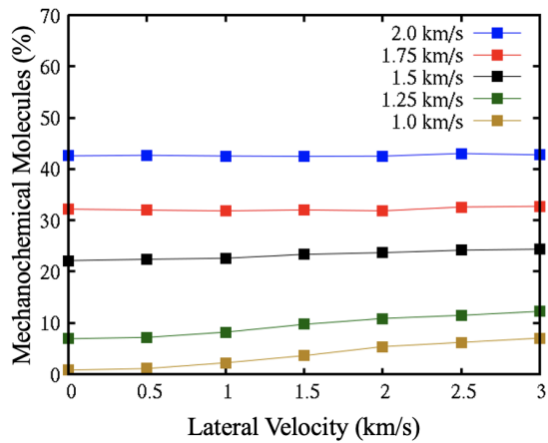

*Figure 10: The percentage of TATB molecules with sufficient $U_{Latent} \geq 100$ kcal/mol required for significant mechanochemical activation (i.e., above pink line in Figure 7b) for the $0°$ case. Each line corresponds to a different compressive velocity. The population analysis considers a 50nm region centered at the interface.*

## 5. PETN

Recent studies on mechanochemical activation of HEs have largely focused on TATB[24,37–39,41,42]. As TATB is a noted outlier in many respects (e.g., its nearly unique safety-performance tradeoffs), and because its unusual chemical reactivity[60] has been tied at least in part to mechanochemical effects[41], it remains an open question whether similar mechanochemical effects arise in other HEs. In this respect, we consider PETN as a model HE without mechanically stiff molecular ring structures to assess the generality of the observations made for TATB (compare molecular structures in Figure 2). PETN crystal-crystal impact simulations are conducted for the same sets of velocities as was done in TATB, using the PETN [001] and [100] directions.

Figure 11 shows snapshots of the molecular deformations at the interfacial hotspot for both PETN orientations for a variety of compressive and lateral velocity conditions, analogous to the results shown for TATB in Figure 5. For both orientations, increasing both compressive and lateral velocity results in a local increase in plasticity and apparent amorphization at the interface. Additionally, the [100] cases appear to exhibit deformation due to increasing lateral velocity at locations further away from the interface as compared to the [001] case. This is especially true for the 2.0 km/s compressive velocity case in which distortions to the lattice exist within the entire region shown. Strongly shocking along [100] activates slip along the $\{110\}\langle 1\bar{1}1 \rangle$ slip system[51,61], which falls at a ±45° angle with respect to the compression direction. Similar slip-mediated "shear bands" have been noted in large-scale MD simulations of PETN shocked along this direction[51].

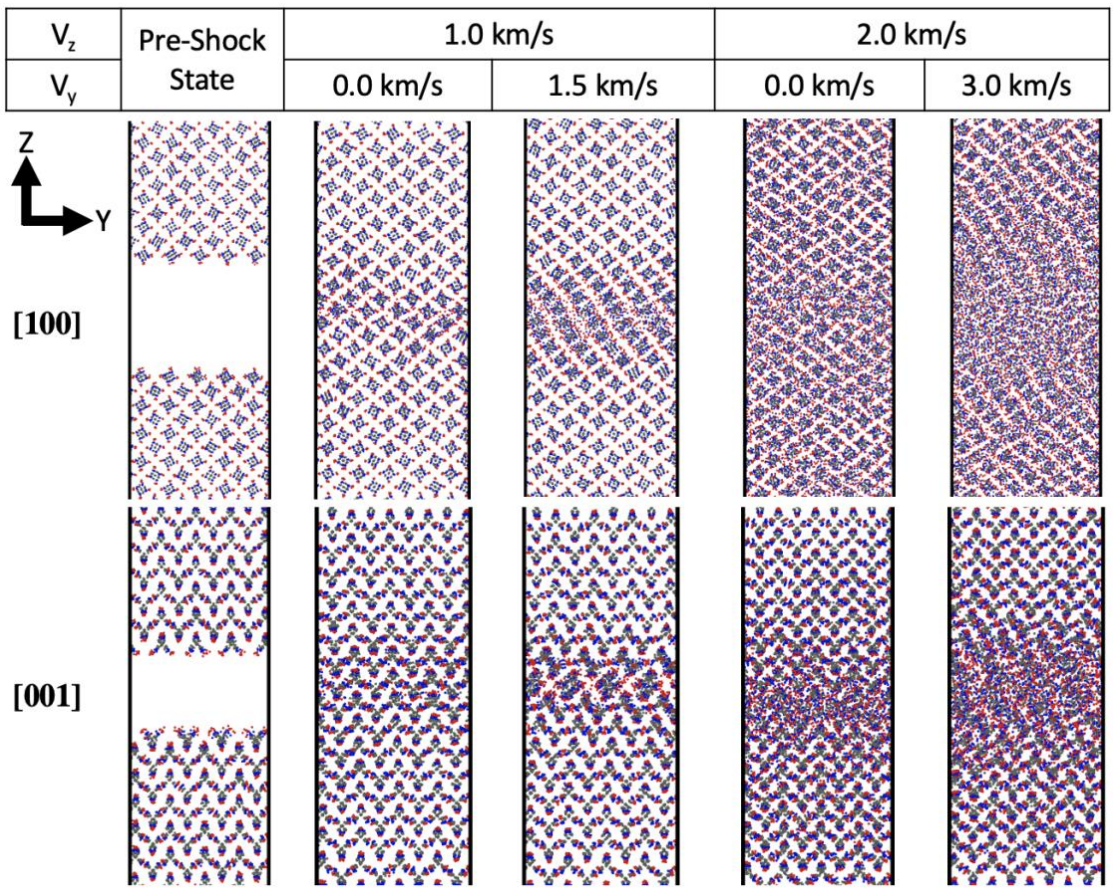

*Figure 11: Snapshots showing PETN packing structure within and around the hotspots formed at crystal-crystal interfaces for selected orientations, compressive velocities, and lateral velocities. Only backbone atoms are plotted (no H or nitro group O) to better visualize local lattice deformations.*

Guided by our previous assessments for TATB in Section 4, we calculated distributions of the per-molecule intra-molecular potential energy and temperature states. Figure 12 shows the resulting $U_{Intra}$-$T$ distributions for the [001] PETN case as parametric functions of compressive and lateral velocity. The PETN results shown here (as well as for the [100] case) are quite similar to the results shown TATB in Figure 7, where increasing compressive velocity increases peak values of both $U_{Intra}$ and $T$ as well as the deviation of $U_{Intra}$ from equipartition. However, increasing lateral velocity only increases the peak values and does not appear to increase the deviation of $U_{Intra}$ from equipartition aside for a few outlier molecules with the largest lateral velocity. This indicates that the intra-molecular strain energy $U_{Latent}$ does not depend strongly on the lateral velocity, which is quite similar to the response seen for TATB.

Figure 13 displays similar $U_{Intra}$-$T$ distributions as Figure 12, but compares the two PETN orientations for the case with 2.0 km/s compressive velocity and 3.0 km/s lateral velocity. Compression along the [100] direction, which is insensitive to shock initiation[56], reaches the same peak values as [001] case and both exhibit similar deviations from equipartition. However, the [001] case has significantly greater density of points at large $T$ and $U_{Intra}$ values. While these results do not indicate a directionally dependent mechanochemical effect, they do indicate that more energy is localized in molecules for the direction known to be more sensitive to initiation. It should be noted that net $U_{Latent}$ of a molecule does not fully determine whether that molecule undergoes mechanochemistry. Recent reactive MD studies indicate that the degree of freedom in the molecule that gains strain energy is a critical factor influencing both chemical dynamics and kinetics[43].

In comparison to TATB, PETN reaches nearly the same quantitative levels of molecular $T$, $U_{Intra}$, and $U_{Latent}$ states in hotspots formed at crystal-crystal interfaces. We also find for both materials that the thermomechanical state of the hotspot and degree of intra-molecular strains are stronger functions of the compressive velocity compared to the lateral shearing velocity. These comparisons indicate that the intra-molecular strain states formed under shock conditions, and their potential for inducing mechanochemical effects, are not just a product of TATB's many unusual physical and chemical characteristics.

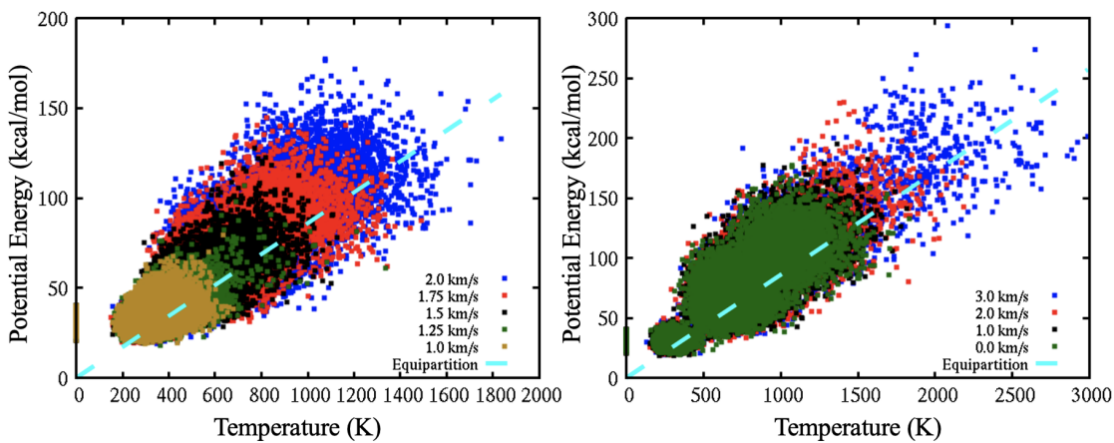

*Figure 12: PE-T ($U_{intra}$ and roto-vibrational T, respectively) distributions for the [001] PETN case. Panel a) shows increasing compressive velocity with no lateral velocity, panel b) shows increasing lateral velocity at 2.0 km/s compressive velocity. Cyan dashed lines represent perfect equipartition of energy.*

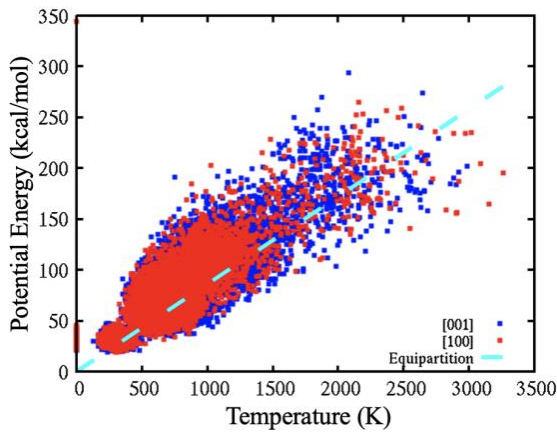

*Figure 13: PE-T ($U_{intra}$ and roto-vibrational T, respectively) distributions showing the effect the compression direction orientation in PETN. Each case is for compressive and lateral velocities of 2.0 km/s and 3.0 km/s, respectively. Cyan dashed lines represent perfect equipartition of energy.*

# 6. Conclusions

We investigated the characteristics of hotspots formed at shocked crystal-crystal interfaces using quasi-1D MD simulations designed to isolate effects due to compression and shear. The generality of the trends identified was surveyed through consideration of two crystalline HE materials (TATB and PETN) with different degrees of molecular conformational flexibility and multiple compression and shearing directions for each material. The simulations contained two crystals separated by a gap. Compressive shocks were generated in the sample through a reverse ballistic approach and a variable lateral velocity component was assigned to one of the crystals, which imposes a shearing/friction effect upon impact. By independently varying both the compressive and shear velocities, as well as material and crystallographic orientation, the resulting effects of interfacial energy localization were assessed in terms of kinetic energy (temperature), potential energy, and intra-molecular strain energy. The intra-molecular strain energy has recently been linked to mechanochemical acceleration of reaction kinetics.

Increasing both compressive and lateral velocity results in an increase in hotspot temperature and potential energy. However, only the compressive work is found to strong effect the intra-molecular strain energy. Assignment of a shearing velocity increases plastic deformation at crystal-crystal interfaces but does not significantly increase intra-molecular strain energy for comparatively strong compressive shocks (P ~ 25 GPa). Shearing does increase intra-molecular strain energy for weaker compressive shocks (P ~ 10 GPa), but the effect is found to be highly localized. By mapping previous reactive MD results onto the intra-molecular strain energy states found here for TATB, we show that regions of significant mechanochemical activity are localized to the crystal-crystal interface. The width of these regions is 4-6 nm for a weak compressive velocity (P ~ 10 GPa) and 10-15nm for a strong compressive velocity (P ~ 25 GPa).

Both TATB and PETN are found to exhibit similar trends regarding the sensitivity of intra-molecular strain energy localization to the compressive and lateral shearing velocity for all crystal orientations considered. Quantitative differences were found for orientations that undergo more extensive plastic deformations, which tend to yield more mechanochemically active molecules at the interface. Overall, this strain energy is shown to be highly correlated with compressive work and considerably less so with shear work. The trends identified here with quasi-1D simulations motivate further elucidation of the convolution of compression and shearing with 2D and 3D interfacial and surface effects, such as shock focusing and jetting.

# Acknowledgements

The authors thank Tommy Sewell and Andrey Pereverzev for providing the LAMMPS implementation of the PETN force field used here.

This work was supported by the U.S. Department of Energy (DOE) through the Los Alamos National Laboratory. The Los Alamos National Laboratory is operated by Triad National Security, LLC, for the National Nuclear Security Administration of the U.S. Department of Energy (Contract No. 89233218CNA000001). Approved for unlimited release: LA-UR-23-21420.

This work was performed under the auspices of the U.S. Department of Energy by Lawrence Livermore National Laboratory under Contract DE-AC52-07NA27344. Work by Purdue University was supported by LLNL subcontract B648789. Approved for unlimited release: LLNL-JRNL-844913-DRAFT.

BWH contributed to this work while at Purdue University and Los Alamos National Laboratory. BWH acknowledges funding provided by the Los Alamos National Laboratory Director's Postdoctoral Fellowship program, project LDRD 20220705PRD1 with partial funding provided by the Advanced Simulation and Computing Physics and Engineering Models project (ASC-PEM).

JM and AS acknowledge funding from the US Office of Naval Research, Multidisciplinary University Research Initiatives (MURI) Program, Contract: N00014-16-1-2557. Program managers: Chad Stoltz and Kenny Lipkowitz.

We acknowledge computational resources from nanoHUB and Purdue University through the Network for Computational Nanotechnology.

# Supplemental Materials to:
# Intergranular Hotspots: A Molecular Dynamics Study on the Influence of Compressive and Shear Work

Brenden W. Hamilton[1,2], Matthew P. Kroonblawd[3], Jalen Macatangay[1], H. Keo Springer[3], and Alejandro Strachan[1]*

Affiliations

[1]School of Materials Engineering and Birck Nanotechnology Center, Purdue University, West Lafayette, Indiana, 47907 USA
[2]Theoretical Division, Los Alamos National Laboratory, Los Alamos, New Mexico 87545, USA
[3]Physical and Life Sciences Directorate, Lawrence Livermore National Laboratory, Livermore, California 94550, USA

* strachan@purdue.edu

# SECTION SM-1

SM Figure 1 shows the baseline rise in $U_{Intra}$ for all materials and orientations considered in the main manuscript plotted as a function of compressive (particle) velocity. Values were averaged over 2nm Eulerian bins in the upstream (right-hand) crystal grain, prior to the shockwave reaching the free surface.

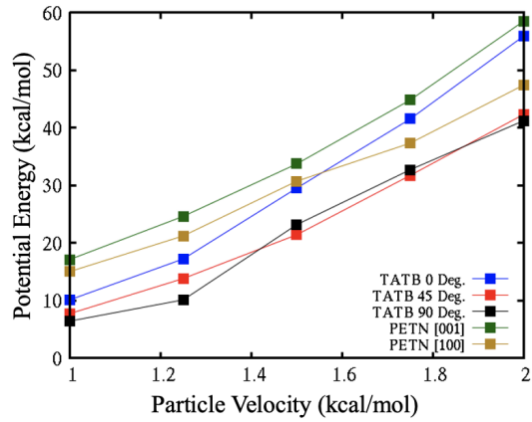

*SM Figure 1: Baseline $U_{Intra}$ measured in the bulk material far from the crystal-crystal interface and free surfaces for all HEs and crystal orientations considered plotted as a function of compressive velocity.*

# SECTION SM-2

SM Figures 2 and 3 mirror main manuscript Figure 8. The main manuscript figure shows results for the TATB 0º case whereas SM 2 and 3 show results for the 45º and 90º cases, respectively. Slices were taken at 5 ps after crystal-crystal impact using 2nm Eulerian bins and mechanochemical groups are based off of reactive MD results.

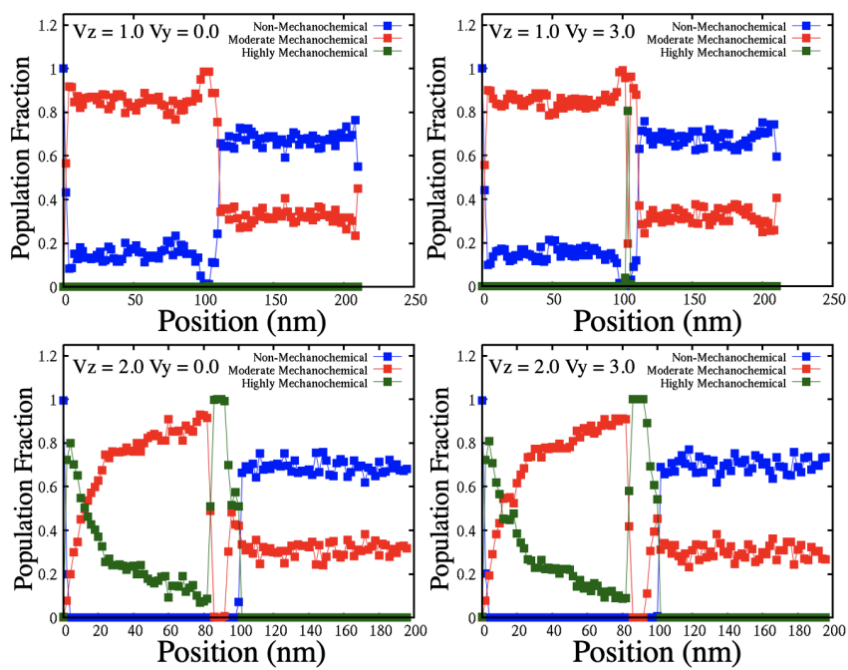

*SM Figure 2: Spatially resolved populations of mechanochemically activated molecules for the 45° case at time $t_o$ + 5 ps. Populations are based on the regions plotted in Figure 7b in the main manuscript, which correspond to the mechanochemical model used in the main manuscript. Population fractions are computed with respect to the total number of molecules in each Eulerian bin. Panels correspond to 1.0 and 2.0 km/s compressive velocity, top and bottom row respectively, and 0.0 and 3.0 km/s lateral velocity, left and right columns, respectively.*

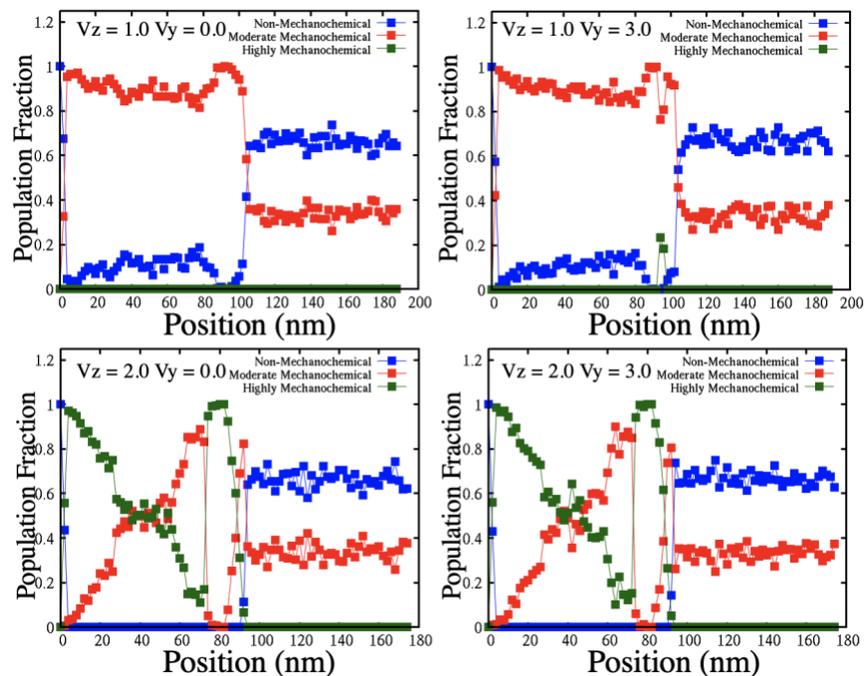

*SM Figure 3: Spatially resolved populations of mechanochemically activated molecules for the 90° case at time $t_o$ + 5 ps. Populations are based on the regions plotted in Figure 7b in the main manuscript, which correspond to the mechanochemical model used in the main manuscript. Population fractions are computed with respect to the total number of molecules in each Eulerian bin. Panels correspond to 1.0 and 2.0 km/s compressive velocity, top and bottom row respectively, and 0.0 and 3.0 km/s lateral velocity, left and right columns, respectively.*

# SECTION SM-3

SM Figure 4 mirrors main manuscript Figure 10. The main manuscript figure shows results for the TATB 0º case whereas SM 4 a and b show results for the 45º and 90º cases, respectively. These show the trends with lateral velocity are consistent across orientations.

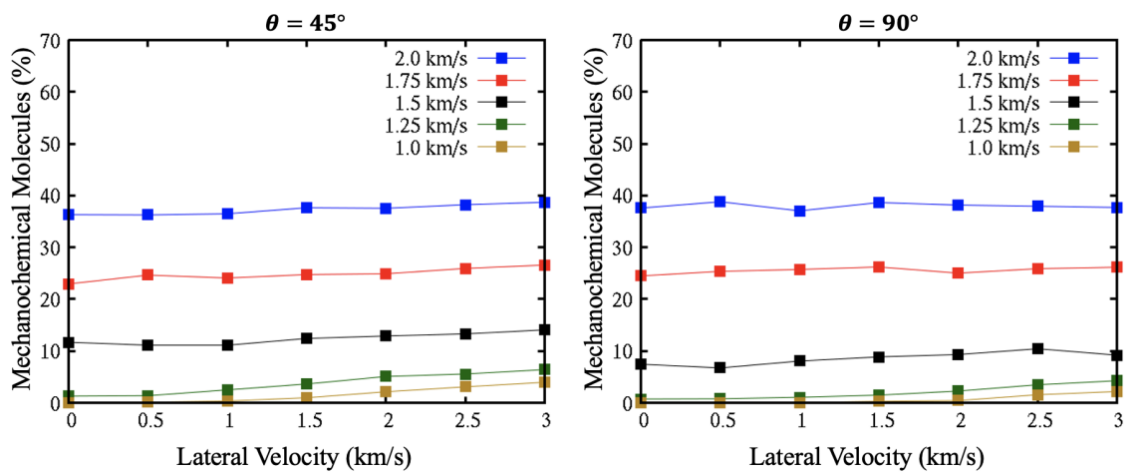

*SM Figure 4: The percentage of molecules with sufficient $U_{Latent} \geq 100$ kcal/mol required for significant mechanochemical activation (i.e., above pink line in Figure 7b in main manuscript) for the 45° and 90º cases. Each line corresponds to a different compressive velocity. The population analysis considers a 50nm region centered at the interface.*

# SECTION SM-4

SM Table 1 shows Pearson Correlation Coefficients for selected quantities averaged in spatial Eulerian bins that were 2nm wide along the compression direction. Only the four nearest bins from either side from the interface were considered in the correlation analysis (4 bins from each of the 105 runs). Quantities Vy and Vz are the initial lateral and compressive velocity, respectively. $W_C$ and $W_T$ are respectively the work done from compression and lateral (shear) motion up to 5 ps after impact. These work components were computed using

$$W = \int_0^t \sigma_{ij} \cdot \frac{dv_j}{dz} \cdot V \cdot dt$$

Here $\upsilon$ is velocity in the j direction, $\sigma_{ij}$ is the stress in the ij tensor term, and $V$ is the bin volume. Quantities $T$ and $U_{Latent}$ are respectively the temperature (rotational-librational KE) and the intra-molecular strain energy, which were computed using the approach described in Section 2 of the main text.

SM Table 1: Pearson Correlation Tables for values of selected material properties obtained in Eulerian bins near the crystal-crystal interface 5ps after impact.

| 0° | Vz | Vy | $W_c$ | $W_T$ | T | $U_{Latent}$ |
|---|---|---|---|---|---|---|
| **Vz** | 1.000 | 0.000 | 0.403 | 0.023 | 0.668 | 0.969 |
| **Vy** | | 1.000 | 0.098 | 0.835 | 0.663 | 0.165 |
| **$W_c$** | | | 1.000 | 0.286 | 0.416 | 0.401 |
| **$W_T$** | | | | 1.000 | 0.733 | 0.145 |
| **T** | | | | | 1.000 | 0.755 |
| **$U_{Latent}$** | | | | | | 1.000 |
| **45°** | **Vz** | **Vy** | **$W_c$** | **$W_T$** | **T** | **$U_{Latent}$** |
| **Vz** | 1.000 | 0.000 | 0.578 | 0.367 | 0.588 | 0.844 |
| **Vy** | | 1.000 | 0.385 | 0.757 | 0.666 | 0.419 |
| **$W_c$** | | | 1.000 | 0.680 | 0.810 | 0.735 |
| **$W_T$** | | | | 1.000 | 0.935 | 0.677 |
| **T** | | | | | 1.000 | 0.844 |
| **$U_{Latent}$** | | | | | | 1.000 |
| **90°** | **Vz** | **Vy** | **$W_c$** | **$W_T$** | **T** | **$U_{Latent}$** |
| **Vz** | 1.000 | 0.000 | 0.878 | 0.442 | 0.868 | 0.957 |
| **Vy** | | 1.000 | 0.187 | 0.657 | 0.330 | 0.148 |
| **$W_c$** | | | 1.000 | 0.697 | 0.953 | 0.874 |
| **$W_T$** | | | | 1.000 | 0.796 | 0.539 |
| **T** | | | | | 1.000 | 0.892 |
| **$U_{Latent}$** | | | | | | 1.000 |

# References

(1) Ravelo, R.; Germann, T. C.; Guerrero, O.; An, Q.; Holian, B. L. Shock-Induced Plasticity in Tantalum Single Crystals: Interatomic Potentials and Large-Scale Molecular-Dynamics Simulations. *Phys. Rev. B - Condens. Matter Mater. Phys.* **2013**, *88* (13), 1–17.

(2) Jaramillo, E.; Sewell, T. D.; Strachan, A. Atomic-Level View of Inelastic Deformation in a Shock Loaded Molecular Crystal. *Phys. Rev. B* **2007**, *76* (6).

(3) Cawkwell, M. J.; Sewell, T. D.; Zheng, L.; Thompson, D. L. Shock-Induced Shear Bands in an Energetic Molecular Crystal: Application of Shock-Front Absorbing Boundary Conditions to Molecular Dynamics Simulations. *Phys. Rev. B.* **2008**, *78* (1), 1–13.

(4) Levitas, V. I.; Ravelo, R. Virtual Melting as a New Mechanism of Stress Relaxation under High Strain Rate Loading. *Proc. Natl. Acad. Sci. U. S. A.* **2012**, *109* (33), 13204–13207.

(5) Macatangay, J.; Hamilton, B. W.; Strachan, A. Deviatoric Stress Driven Transient Melting Below the Glass Transition Temperature in Shocked Polymers. *J. Appl. Phys.* **2022**, *132* (3), 035901.

(6) Bourne, N.; Millett, J.; Rosenberg, Z.; Murray, N. On the Shock Induced Failure of Brittle Solids. *J. Mech. Phys. Solids* **1998**, *46* (10), 1887–1908.

(7) Grilli, N.; Duarte, C. A.; Koslowski, M. Dynamic Fracture and Hot-Spot Modeling in Energetic Composites. *J. Appl. Phys.* **2018**, *123* (6).

(8) Hamilton, B. W.; Sakano, M. N.; Li, C.; Strachan, A. Chemistry Under Shock Conditions. *Annu. Rev. Mater. Res.* **2021**, *51* (1), 101–130.

(9) Brown, K. E.; McGrane, S. D.; Bolme, C. A.; Moore, D. S. Ultrafast Chemical Reactions in Shocked Nitromethane Probed with Dynamic Ellipsometry and Transient Absorption Spectroscopy. *J. Phys. Chem. A* **2014**, *118* (14), 2559–2567.

(10) Strachan, A.; van Duin, A. C. T.; Chakraborty, D.; Dasgupta, S.; Goddard, W. A. Shock Waves in High-Energy Materials: The Initial Chemical Events in Nitramine RDX. *Phys. Rev. Lett.* **2003**, *91* (9), 7–10.

(11) Handley, C. A.; Lambourn, B. D.; Whitworth, N. J.; James, H. R.; Belfield, W. J. Understanding the Shock and Detonation Response of High Explosives at the Continuum and Meso Scales. *Appl. Phys. Rev.* **2018**, *5* (1), 011303.

(12) Davis, W. C. High Explosives The Interaction of Chemistry and Mechanics. *Los Alamos Sci.* **1981**, *2* (1), 48–75.

(13) Campbell, A. W.; Travis, J. R. The Shock Desensitization of Pbx-9404 and Composition B-3. *Los Alamos Natl. Lab.* **1985**, No. LA-UR-85-114.

(14) Austin, R. A.; Barton, N. R.; Reaugh, J. E.; Fried, L. E. Direct Numerical Simulation of Shear Localization and Decomposition Reactions in Shock-Loaded HMX Crystal. *J. Appl. Phys.* **2015**, *117* (18).

(15) Springer, H. K.; Bastea, S.; Nichols, A. L.; Tarver, C. M.; Reaugh, J. E. Modeling The Effects of Shock Pressure and Pore Morphology on Hot Spot Mechanisms in HMX. *Propellants, Explos. Pyrotech.* **2018**, *43* (8), 805–817.

(16) Kapahi, A.; Udaykumar, H. S. Dynamics of Void Collapse in Shocked Energetic Materials: Physics of Void-Void Interactions. *Shock Waves* **2013**, *23* (6), 537–558.

(17) Rai, N. K.; Udaykumar, H. S. Void Collapse Generated Meso-Scale Energy Localization in Shocked Energetic Materials: Non-Dimensional Parameters, Regimes, and Criticality of Hotspots. *Phys. Fluids* **2019**, *31* (1).

(18) Zhao, P.; Lee, S.; Sewell, T.; Udaykumar, H. S. Tandem Molecular Dynamics and Continuum Studies of Shock-Induced Pore Collapse in TATB. *Propellants, Explos.*

*Pyrotech.* **2020**, *45* (2), 196–222.
(19) Duarte, C. A.; Li, C.; Hamilton, B. W.; Strachan, A.; Koslowski, M. Continuum and Molecular Dynamics Simulations of Pore Collapse in Shocked β -Tetramethylene Tetranitramine ( β -HMX) Single Crystals. *J. Appl. Phys.* **2021**, *129* (1), 015904.
(20) Wood, M. A.; Kittell, D. E.; Yarrington, C. D.; Thompson, A. P. Multiscale Modeling of Shock Wave Localization in Porous Energetic Material. *Phys. Rev. B* **2018**, *97* (1), 1–9.
(21) Miller, C.; Olsen, D.; Wei, Y.; Zhou, M. Three-Dimensional Microstructure-Explicit and Void-Explicit Mesoscale Simulations of Detonation of HMX at Millimeter Sample Size Scale. *J. Appl. Phys.* **2020**, *127* (12), 125105.
(22) Kroonblawd, M. P.; Hamilton, B. W.; Strachan, A. Fourier-like Thermal Relaxation of Nanoscale Explosive Hot Spots. *J. Phys. Chem. C* **2021**, *125*, 20570–20582.
(23) Li, C.; Hamilton, B. W.; Shen, T.; Alzate, L.; Strachan, A. Systematic Builder for All-Atom Simulations of Plastically Bonded Explosives. *Propellants, Explos. Pyrotech.* **2022**, 47 (8), e202200003.
(24) Hamilton, B. W.; Germann, T. C. Energy Localization Efficiency in 1,3,5-Trinitro-2,4,6-Triaminobenzene Pore Collapse Mechanisms. *J. Appl. Phys.* **2023**, *133*, 035901.
(25) Holian, B. L.; Germann, T. C.; Maillet, J. B.; White, C. T. Atomistic Mechanism for Hot Spot Initiation. *Phys. Rev. Lett.* **2002**, *89* (28), 1–4.
(26) Li, C.; Hamilton, B. W.; Strachan, A. Hotspot Formation Due to Shock-Induced Pore Collapse in 1,3,5,7-Tetranitro-1,3,5,7-Tetrazoctane (HMX): Role of Pore Shape and Shock Strength in Collapse Mechanism and Temperature. *J. Appl. Phys.* **2020**, *127* (17), 175902.
(27) Cates, J. E.; Sturtevant, B. Shock Wave Focusing Using Geometrical Shock Dynamics. *Phys. Fluids* **1997**, *9* (10), 3058–3068.
(28) Sturtevant, B.; Kulkarny, V. A. The Focusing of Weak Shock Waves. *J. Fluid Mech.* **1976**, *73* (4), 651–671.
(29) Wiita, A. P.; Ainavarapu, S. R. K.; Huang, H. H.; Fernandez, J. M. Force-Dependent Chemical Kinetics of Disulfide Bond Reduction Observed with Single-Molecule Techniques. *Proc. Natl. Acad. Sci. U. S. A.* **2006**, *103* (19), 7222–7227.
(30) Piermattei, A.; Karthikeyan, S.; Sijbesma, R. P. Activating Catalysts with Mechanical Force. *Nat. Chem.* **2009**, *1* (2), 133–137.
(31) Kingsbury, C. M.; May, P. A.; Davis, D. A.; White, S. R.; Moore, J. S.; Sottos, N. R. Shear Activation of Mechanophore-Crosslinked Polymers. *J. Mater. Chem.* **2011**, *21* (23), 8381–8388.
(32) Zhou, X.; Miao, Y.; Suslick, K. S.; Dlott, D. D. Mechanochemistry of Metal-Organic Frameworks under Pressure and Shock. *Acc. Chem. Res.* **2020**, *53* (12).
(33) Wang, J.; Kouznetsova, T. B.; Niu, Z.; Ong, M. T.; Klukovich, H. M.; Rheingold, A. L.; Martinez, T. J.; Craig, S. L. Inducing and Quantifying Forbidden Reactivity with Single-Molecule Polymer Mechanochemistry. *Nat. Chem.* **2015**, *7* (4), 323–327.
(34) Davis, D. A.; Hamilton, A.; Yang, J.; Cremar, L. D.; Van Gough, D.; Potisek, S. L.; Ong, M. T.; Braun, P. V.; Martínez, T. J.; White, S. R.; Moore, J. S.; Sottos, N. R. Force-Induced Activation of Covalent Bonds in Mechanoresponsive Polymeric Materials. *Nature* **2009**, *459* (7243), 68–72.
(35) Wood, M. A.; Cherukara, M. J.; Kober, E. M.; Strachan, A. Ultrafast Chemistry under Nonequilibrium Conditions and the Shock to Deflagration Transition at the Nanoscale. *J. Phys. Chem. C* **2015**, *119* (38), 22008–22015.

(36) Islam, M. M.; Strachan, A. Role of Dynamical Compressive and Shear Loading on Hotspot Criticality in RDX via Reactive Molecular Dynamics. *J. Appl. Phys.* **2020**, *128* (6).

(37) Hamilton, B. W.; Kroonblawd, M. P.; Li, C.; Strachan, A. A Hotspot's Better Half: Non-Equilibrium Intra-Molecular Strain in Shock Physics. *J. Phys. Chem. Lett.* **2021**, *12* (11), 2756–2762.

(38) Hamilton, B. W.; Kroonblawd, M. P.; Strachan, A. The Potential Energy Hotspot: Effects of Impact Velocity, Defect Geometry, and Crystallographic Orientation. *J. Phys. Chem. C* **2022**, *126* (7), 3743-3755

(39) Hamilton, B. W.; Kroonblawd, M. P.; Strachan, A. Extemporaneous Mechanochemistry: Shockwave Induced Ultrafast Chemical Reactions Due to Intramolecular Strain Energy. *J. Phys. Chem. Lett.* **2022**, *13*, 6657–6663.

(40) Hamilton, B. W.; Germann, T. C. Interplay of Mechanochemistry and Material Processes in the Graphite to Diamond Phase Transformation. *arXiv* **2023**.

(41) Kroonblawd, M. P.; Fried, L. E. High Explosive Ignition through Chemically Activated Nanoscale Shear Bands. *Phys. Rev. Lett.* **2020**, *124* (20), 206002.

(42) Kroonblawd, M. P.; Steele, B. A.; Nelms, M. D.; Fried, L. E.; Austin, R. A. Anisotropic Strength Behavior of Single-Crystal TATB. *Model. Simul. Mater. Sci. Eng.* **2021**, *30* (1), 014004.

(43) Hamilton, B. W.; Strachan, A. Many-Body Mechanochemistry : Intra-Molecular Strain in Condensed Matter Chemistry. *chemRxiv* **2022**.

(44) Plimpton, S. Fast Parallel Algorithms for Short-Range Molecular Dynamics. *J. Comput. Phys.* **1995**, *117* (1), 1–19.

(45) Thompson, A. P.; Aktulga, H. M.; Berger, R.; Bolintineanu, D. S.; Brown, W. M.; Crozier, P. S.; in 't Veld, P. J.; Kohlmeyer, A.; Moore, S. G.; Nguyen, T. D.; Shan, R.; Stevens, M. J.; Tranchida, J.; Trott, C.; Plimpton, S. J. LAMMPS - a Flexible Simulation Tool for Particle-Based Materials Modeling at the Atomic, Meso, and Continuum Scales. *Comput. Phys. Commun.* **2022**, *271*, 108171.

(46) Bedrov, D.; Borodin, O.; Smith, G. D.; Sewell, T. D.; Dattelbaum, D. M.; Stevens, L. L. A Molecular Dynamics Simulation Study of Crystalline 1,3,5-Triamino-2,4,6-Trinitrobenzene as a Function of Pressure and Temperature. *J. Chem. Phys.* **2009**, *131* (22).

(47) Kroonblawd, M. P.; Sewell, T. D. Theoretical Determination of Anisotropic Thermal Conductivity for Initially Defect-Free and Defective TATB Single Crystals. *J. Chem. Phys.* **2014**, *141* (18).

(48) Mathew, N.; Sewell, T. D.; Thompson, D. L. Anisotropy in Surface-Initiated Melting of the Triclinic Molecular Crystal 1,3,5-Triamino-2,4,6-Trinitrobenzene: A Molecular Dynamics Study. *J. Chem. Phys.* **2015**, *143* (9).

(49) Wolf, D.; Keblinski, P.; Phillpot, S. R.; Eggebrecht, J. Exact Method for the Simulation of Coulombic Systems by Spherically Truncated, Pairwise r-1 Summation. *J. Chem. Phys.* **1999**, *110* (17), 8254–8282.

(50) Borodin, O.; Smith, G. D.; Sewell, T. D.; Bedrov, D. Polarizable and Nonpolarizable Force Fields for Alkyl Nitrates. *J. Phys. Chem. B* **2008**, *112* (3), 734–742.

(51) Eason, R. M.; Sewell, T. D. Shock-Induced Inelastic Deformation in Oriented Crystalline Pentaerythritol Tetranitrate. *J. Phys. Chem. C* **2012**, *116* (3), 2226–2239.

(52) Pollock, E. L.; Glosli, J. Comments on P3M, FMM, and the Ewald Method for Large

Periodic Coulombic Systems. *Comput. Phys. Commun.* **1996**, *95* (2–3), 93–110.

(53) Nosé, S. A Unified Formulation of the Constant Temperature Molecular Dynamics Methods. *J. Chem. Phys.* **1984**, *81* (1), 511–519. https://doi.org/10.1063/1.447334.

(54) Holian, B. L.; Straub, G. K. Molecular Dynamics of Shock Waves in Three-Dimensional Solids: Transition from Nonsteady to Steady Waves in Perfect Crystals and Implications for the Rankine-Hugoniot Conditions. *Phys. Rev. Lett.* **1979**, *43* (21), 1598–1600.

(55) Kroonblawd, M. P.; Mathew, N.; Jiang, S.; Sewell, T. D. A Generalized Crystal-Cutting Method for Modeling Arbitrarily Oriented Crystals in 3D Periodic Simulation Cells with Applications to Crystal–Crystal Interfaces. *Comput. Phys. Commun.* **2016**, *207*, 232–242.

(56) Gruzdkov, Y. A.; Gupta, Y. M. Shock Wave Initiation of Pentaerythritol Tetranitrate Single Crystals: Mechanism of Anisotropic Sensitivity. *J. Phys. Chem. A* **2000**, *104* (47), 11169–11176.

(57) Shan, T. R.; Wixom, R. R.; Mattsson, A. E.; Thompson, A. P. Atomistic Simulation of Orientation Dependence in Shock-Induced Initiation of Pentaerythritol Tetranitrate. *J. Phys. Chem. B* **2013**, *117* (3), 928–936.

(58) Lafourcade, P.; Denoual, C.; Maillet, J. B. Irreversible Deformation Mechanisms for 1,3,5-Triamino-2,4,6-Trinitrobenzene Single Crystal through Molecular Dynamics Simulations. *J. Phys. Chem. C* **2018**, *122* (26), 14954–14964.

(59) Zhao, P.; Kroonblawd, M. P.; Mathew, N.; Sewell, T. Strongly Anisotropic Thermomechanical Response to Shock Wave Loading in Oriented Samples of the Triclinic Molecular Crystal 1,3,5- Triamino-2,4,6-Trinitrobenzene. *J. Phys. Chem. C* **2021**, *125*, 22747–22765.

(60) Wescott, B. L.; Stewart, D. S.; Davis, W. C. Equation of State and Reaction Rate for Condensed-Phase Explosives. *J. Appl. Phys.* **2005**, *98* (5).

(61) Dick, J. J.; Ritchie, J. P. Molecular Mechanics Modeling of Shear and the Crystal Orientation Dependence of the Elastic Precursor Shock Strength in Pentaerythritol Tetranitrate. *J. Appl. Phys.* **1994**, *76* (5), 2726–2737.